# High-Efficiency Wakefield Acceleration of Positrons via a Slit Plasma Channel

Yu-Jie Wang[1], Xing-Long Zhu[1*], Hong-Kang Zuo[1], and Zheng-Ming Sheng[2, 3]

[1] Institute for Fusion Theory and Simulation, School of Physics, Zhejiang University, Hangzhou 310058, China

[2] State Key Laboratory of Dark Matter Physics, Key Laboratory for Laser Plasmas (MOE), School of Physics and Astronomy, Shanghai Jiao Tong University, Shanghai 200240, China

[3] Tsung-Dao Lee Institute, Shanghai Jiao Tong University, Shanghai 201210, China

[*]Email: xinglong.zhu@zju.edu.cn (X.L.Z.)

**Abstract:** Although plasma-based wakefield acceleration has achieved remarkable success for electrons, efficient positron acceleration remains a long-standing challenge, as positrons suffer severe defocusing in the accelerating phases of such structures. Here, we propose a slit plasma wakefield acceleration (SPWA) scheme, in which a relativistic electron beam propagating through a slit channel formed by two parallel plasma slabs excites crossing plasma-electron filamentary structures within the slit, generating intense wakefields tailored to simultaneously provide longitudinal acceleration and transverse focusing for a trailing positron bunch. A theoretical model is developed to characterize the wakefield structure responsible for positron acceleration and focusing. Three-dimensional particle-in-cell simulations validate the model, and demonstrate that the SPWA scheme achieves a positron trapping efficiency of nearly 100%, while maintaining low emittance growth and narrow energy spread for an injected positron beam. These findings establish a new and robust approach to high-efficiency positron acceleration, with promising implications for future compact electron-positron colliders and high-energy positron applications.

Plasma-based accelerators have revolutionized high-energy particle acceleration, providing accelerating gradients more than three orders of magnitude higher than those achievable with conventional radio-frequency technology, and thus offering transformative potential for downsizing and cost reduction of future particle colliders [1-3]. Since the pioneering proposals of laser- and particle-beam-driven plasma wakefield acceleration [4, 5], remarkable progress has been made for relativistic electron acceleration, including extremely high acceleration gradients with multi-GeV energy gain and well-controlled beam quality with narrow energy spread and low emittance [6-10]. However, this success has not been paralleled in the domain of positron acceleration, where achieving comparable performance remains a persistent challenge. This asymmetry poses a critical bottleneck for the development of plasma-based electron-positron colliders, which are being considered for precision Higgs physics and the search for new physics beyond the Standard Model [11-13].

It is well-known that the plasma wakefields driven either by an intense laser pulse or a particle beam can provide both strong longitudinal acceleration fields and transverse focusing forces for trapped electrons [14-16]. For positrons, however, the transverse fields of such wakes usually defocus particles except in an extremely narrow region at the rear of the bubble where the returning plasma electrons cross the axis [17, 18]. This structure severely limits positron trapping, beam quality and acceleration efficiency. Significant efforts have been made to address this issue. For example, hollow plasma channels have been widely investigated as an approach to sustain high acceleration gradients while mitigating transverse defocusing [19-23]. However, such a configuration is typically highly susceptible to beam breakup instabilities triggered by even tiny misalignments between the beam and the channel axis [24], leading to rapid emittance growth and substantial particle loss. Alternative approaches, such as donut-shaped drivers [25-27], beam-driven self-loaded wakefields [28], as well as some specially structured plasma targets [29, 30], have been explored to form a field structure suitable both for focusing and acceleration of positrons. Nevertheless, most of these schemes are hindered by limited energy transfer efficiency, low accelerated charge, and/or transverse instabilities. In addition, stringent requirements on beam alignment and matching pose substantial experimental challenges, severely restricting their practical applicability. There is thus an urgent need for a robust and feasible scheme capable of generating stable wakefields for high-efficiency positron acceleration without relying on complex drivers or prohibitive technical requirements.

In this Letter, we propose a scheme termed slit-plasma wakefield acceleration (SPWA), in which a distinctive acceleration structure is formed as an ultra-relativistic electron beam driver propagates through a slit plasma channel, enabling the simultaneous acceleration and focusing of trapped positrons. This favorable field configuration permits efficient loading of high-charge positron bunches and supports their stable propagation over extended acceleration distances. Consequently, positrons can be accelerated to high energies while preserving low emittance and narrow energy spread.

We first introduce the basic idea of the scheme. As schematically illustrated in Fig. 1, it employs a slit-plasma channel formed by two parallel plasma slabs separated by a narrow vacuum gap, through which an ultra-relativistic electron beam propagates. As the electron beam travels through the slit channel, plasma electrons are expelled from the inner walls of the target, leaving behind the massive ions. These exposed ions pull the expelled plasma electrons back, forming thin electron sheaths. The returning sheath electrons overshoot the initial plasma boundaries and flow into the central region of the slit. This dynamic process excites large-amplitude plasma wakes and forms symmetric half-bubble cavities on both sides of the slit, as shown in Figs. 1(a) and 1(c). The excited wakefields support high acceleration gradients in the longitudinal direction while simultaneously providing strong transverse focusing forces that enable stable confinement and propagation of trailing positrons. These features clearly distinguish our scheme from the widely studied approach based on hollow plasma channels, which often lacks robust transverse focusing and is susceptible to severe beam-breakup instabilities.

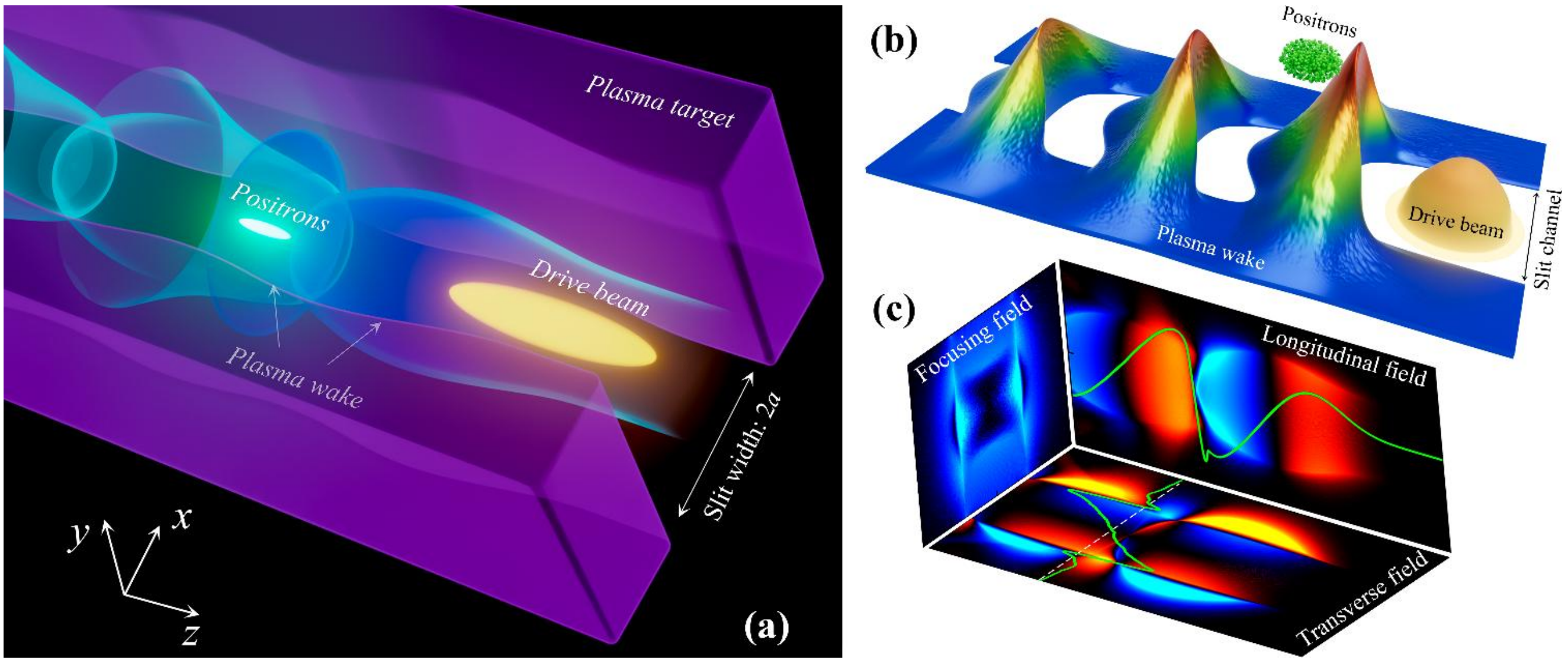

**Fig. 1.** (a) Schematic of positron acceleration in a slit plasma channel. (b) The plasma wake is driven by an electron beam passing through the slit plasma channel and a trailing positron beam is injected at the suitable phase for acceleration. (c) A region suitable for acceleration and focusing of trailing positrons is identified just behind the first wake bucket.

To demonstrate the proposed scheme, we start by showing simulation results obtained with particle-in-cell (PIC) codes EPOCH [31] and SMILEI [32], which yield consistent results. In the three-dimensional (3D) PIC simulations, the window has dimensions of $L_x \times L_y \times L_z = 130\mu\text{m} \times 130\mu\text{m} \times 240\mu\text{m}$ with grid cells of $650 \times 650 \times 1440$ and 2 macroparticles per cell for all species. The window moves along the $z$-axis at the speed $c$ of light in vacuum. For simplicity, the slit-shaped plasma target is set to have a step density profile along a transverse $x$-direction with

$$n_{pe}(x) = \begin{cases} 0, & |x| < a \\ n_{e0}, & a \leq |x| \end{cases} \tag{1}$$

where $n_{e0} = 4 \times 10^{17}\text{cm}^{-3}$ is the plasma density, and $a = 25\mu\text{m}$ represents the distance from the center of the slit to the target. The plasma and beam parameters can all be adjusted, as shown in the following parameter study.

Figure 2 illustrates the formation of the tailored wakefields in the slit plasma. As an example, we adopt a 5 GeV electron beam as the driver. The beam has a total charge of $Q_b = 2\text{nC}$, with Gaussian spatial and momentum distributions characterized by $\sigma_z = 18\mu\text{m}$ and $\sigma_x = \sigma_y = 8\mu\text{m}$, rms energy spread 0.43%, and normalized emittance $\varepsilon_n = 16.6\ \text{mm} \cdot \text{mrad}$. The driving beam rapidly expels ambient plasma electrons outward from the inner surfaces of the target, while the massive background ions remain nearly immobile. These exposed ions then pull the expelled plasma electrons back toward the slit center, forming thin electron sheaths that bound the developing wake bucket. The returning sheath electrons overshoot their initial plasma boundaries, forming a distinct crossed-electron trajectory pattern [Figs. 2(a-c)], which reshapes and dominates the transverse wakefield inside the slit channel. Unlike the ion-rich environment of usual nonlinear wakes, this slit geometry establishes an extended ion-free central region populated by dense plasma electrons. This charge distribution generates a strong longitudinal acceleration field and a transverse focusing field, making it suitable for simultaneously confining and accelerating positively charged positrons, as exhibited in Figs. 2(d-i). During this process, a quadrupole-like transverse wakefield can be excited across the slit [Figs. 2(j1-$l$1)], which is described by $\vec{W}_\perp = (E_x - cB_y)\hat{x} + (E_y + cB_x)\hat{y}$. This field induces anisotropic beam evolution: the driving beam experiences defocusing along the $x$-direction while being simultaneously focused along the $y$-direction until most of the beam electrons reach the target walls. In conventional radio-frequency and dielectric accelerators, analogous high-order wakefield effects may arise, yet such wakefields must be actively suppressed or precisely corrected [33]; otherwise, the particles striking the chamber walls are irreversibly lost. The slit-plasma configuration, by contrast, offers a fundamentally different stabilization mechanism. Instead of being immediately lost upon reaching the boundary, beam particles interact with the slit plasma, expelling the local plasma electrons and exposing the background ions. Subsequently, a stable dynamic equilibrium is achieved, as shown in Fig. 2(c). In this situation, the defocusing forces can be counterbalanced by the strong Coulomb restoring forces exerted by the exposed plasma ions, thereby preventing the beam particles from continuously deviating towards the boundary.

As visualized in the beam evolution presented in Figs. 2(j2-$l$2) for the transverse distributions of the beam density, the beam gradually expands along the $x$-direction while compressing along the $y$-direction. The beam is dense enough to evacuate the ambient plasma electrons and expose the background ions, yet it does not penetrate through the target wall. This is because the strong Coulomb field from these exposed ions exerts a restoring force that pulls the beam electrons back. During the interaction, a distinct beam splitting process appears: the beam electrons initially located at the center of the slit channel are driven toward the plasma walls by the slit-induced quadrupole-like transverse wake, causing the beam to split. After propagation over a distance at the centimeter scale, most of the beam electrons become trapped close to the inner walls, eventually settling into a quasi-steady state. At equilibrium, the initial single beam evenly splits into two symmetric stable beam lobes that maintain steady long-distance propagation without beam electron loss. The wakefields generated by such a modulated beam provide both longitudinal accelerating fields and net transverse focusing forces for trailing positrons.

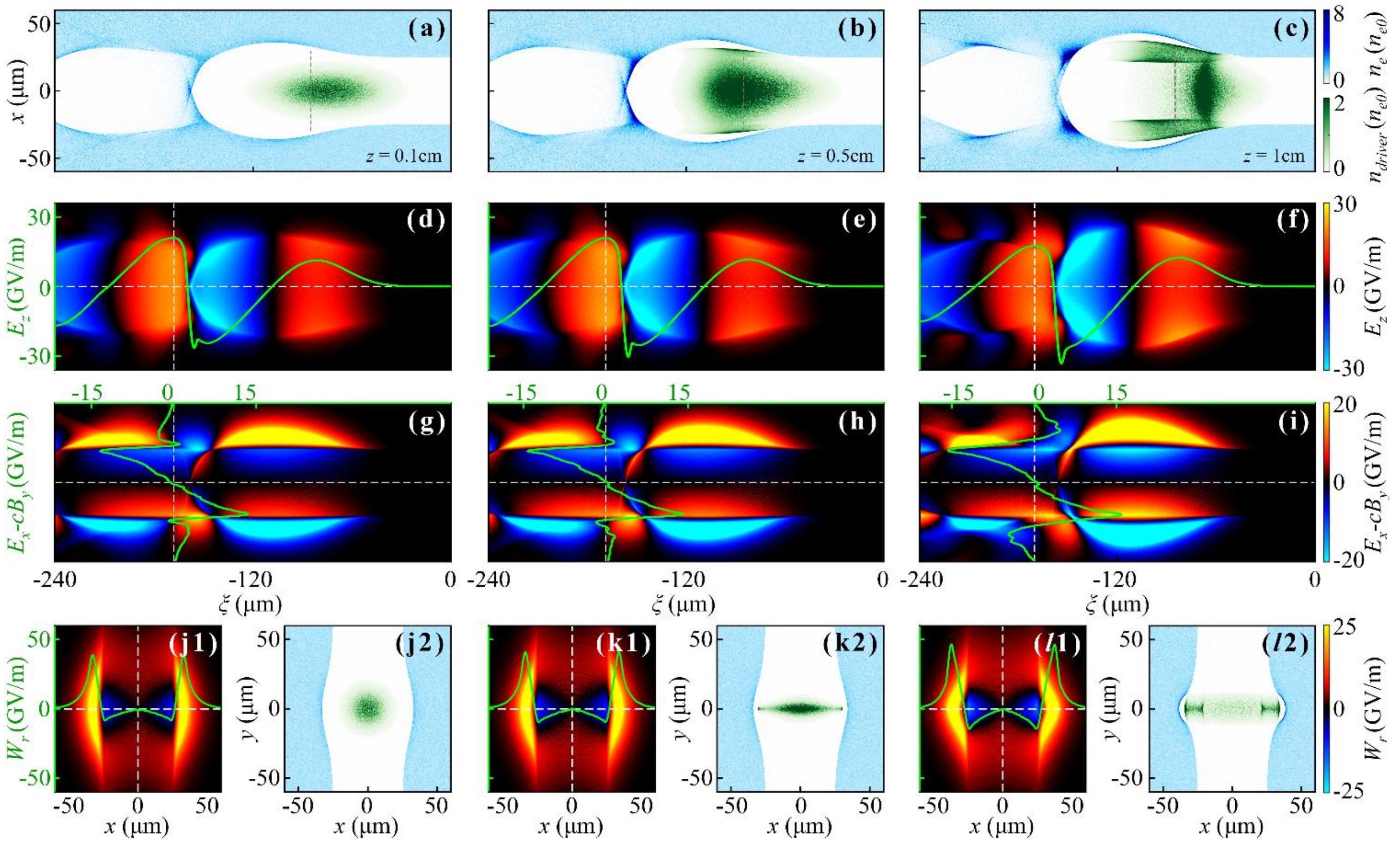

**Fig. 2.** Snapshots showing the evolution of the driving beam and its excited wakefields in the slit plasma. (a, b, c) Beam and plasma densities. (d, e, f) Longitudinal wakefields. (g, h, i) Transverse wakefields. (j1, k1, *l*1) and (j2, k2, *l*2) show respectively the transverse distributions of the transverse field $W_r = \overrightarrow{W}_\perp \cdot \hat{r}$ and the beam and plasma densities at $\xi = -85$μm, where $\hat{r}$ is the radial unit vector in the transverse plane, and $\xi = z - ct$.

To describe the slit-plasma wake and its characteristics found in the simulations shown above, we present an analytical model. A cylindrical hollow channel supports discrete azimuthal modes [20], whereas translational invariance along the slit opening gives a continuous spectrum indexed by the transverse wavenumber $k_y$, and the resulting physical process is different. The corresponding eigen-mode fields determine the longitudinal wake and the transverse response responsible for the driver splitting. We adopt the quasistatic approximation and use the comoving coordinate $\xi = z - ct$. For each $k_y$, the longitudinal eigenfield satisfies $(\partial_x^2 - k_y^2)E_z = 0$ in the vacuum slit and $(\partial_x^2 - k_y^2 - k_p^2)E_z = 0$ in the plasma slabs, where $k_p = \omega_p/c$, and $\omega_p = (n_{e0}e^2/\epsilon_0 m_e)^{1/2}$. The plasma solution decays away from the walls as $\exp\left[-\sqrt{k_y^2 + k_p^2}(|x| - a)\right]$. The eigenmodes separate into even- and odd-$x$ branches according to the parity of $E_z$ about $x = 0$. A centered driver symmetric in both transverse directions excites only the even branch, with $\Phi_{k_y}(x, y) = \cosh(k_y x)\cos(k_y y)$. Matching the tangential fields $E_y$, $E_z$, $B_y$, and $B_z$ at $x = \pm a$ gives [34]

$$\Omega_s^2(k_y) = \frac{k_y}{k_y + \sqrt{k_y^2 + k_p^2}\tanh(k_y a)}, \text{ and } \Omega_s(0) \equiv \lim_{k_y \to 0} \Omega_s(k_y) = \frac{1}{\sqrt{1 + k_p a}}. \quad (2)$$

Here $\Omega_s(k_y)$ is the eigenfrequency of the even-$x$ branch normalized to the plasma frequency $\omega_p$. Figure 3(a) tests its $k_y = 0$ limit using PIC simulations with $Q_b = 0.2$nC and $15 \le a \le 30$μm, keeping the remaining parameters unchanged from Fig. 2. Averaging $E_z$ along $y$ selects this component. The frequency extracted before appreciable beam splitting follows the predicted decrease with increasing slit half-width.

The wake of a driver with finite extent along $y$ is obtained by integrating over the full $k_y$ spectrum. The spectral loss-factor density $\kappa_s(k_y)$ quantifies energy transfer from the driver to each spectral component and determines the wake amplitude [34]. Let $\lambda(\xi) \le 0$ denote the signed electron line-charge density, with $\int \lambda d\xi = -Q_b$. For a separable beam charge density $\rho_b(x, y, \xi) = \lambda(\xi) f_\perp(x, y)$, $f_\perp$ is the normalized transverse profile satisfying $\int f_\perp dxdy = 1$. The excitation of each mode is weighted by $I(k_y) = \int f_\perp(x', y')\Phi_{k_y}(x', y')dx'dy'$. Extending the vacuum eigenfunction over the Gaussian transverse distribution gives $I(k_y) = \exp\left[k_y^2(\sigma_x^2 - \sigma_y^2)/2\right]$, which equals unity for the round driver with $\sigma_x = \sigma_y$. Convolution with $\lambda(\xi)$ then yields the longitudinal field:

$$E_z(x, y, \xi) = -\frac{1}{\pi}\int_0^\infty \kappa_s(k_y) I(k_y) \Phi_{k_y}(x, y) \times \int_\xi^\infty \lambda(\xi')\cos\left[\Omega_s(k_y)k_p(\xi' - \xi)\right] d\xi' dk_y. \quad (3)$$

Only preceding source slices, $\xi' > \xi$, contribute. For a round Gaussian driver centered at $\xi_b$ with rms length $\sigma_z$, maximizing the Gaussian-convolved spectrum over the trailing positron-accelerating bucket gives [34]:

$$E_{acc}^{peak} \simeq \frac{Q_b e^2 n_{e0}}{\pi \epsilon_0^2 m_e c^2}\mathcal{F}_{pk}(k_p a, k_p \sigma_z). \quad (4)$$

Here $\mathcal{F}_{pk}(k_p a, k_p \sigma_z) = \left\{\int_0^\infty \frac{\exp\left[-\frac{\Omega_s^2(k_y)k_p^2\sigma_z^2}{2}\right]\cos\left[\Omega_s(k_y)k_p(\xi_b - \xi)\right]}{2\sqrt{k_y^2 + k_p^2}\cosh(2k_y a) + \left(2k_y + \frac{k_p^2}{k_y}\right)\sinh(2k_y a)} dk_y\right\}_{max}$ is the dimensionless peak factor. This indicates that for a fixed plasma density and beam parameters, the longitudinal field depends on the half-width of the slit structure. Figure 3(b) compares the prediction with simulation results over $0 \le a \le 40$μm, with the remaining parameters as those in Fig. 2. The predicted field $E_{acc}^{peak}$ decreases with increasing $a$ and agrees with PIC simulations over $5 \le a \le 40\mu\text{m}$. For $a \ll \sigma_x = 8\mu\text{m}$, substantial overlap between the driver and the plasma excites a nonlinear blowout response, where the point for $a = 0$ provides the uniform-plasma reference.

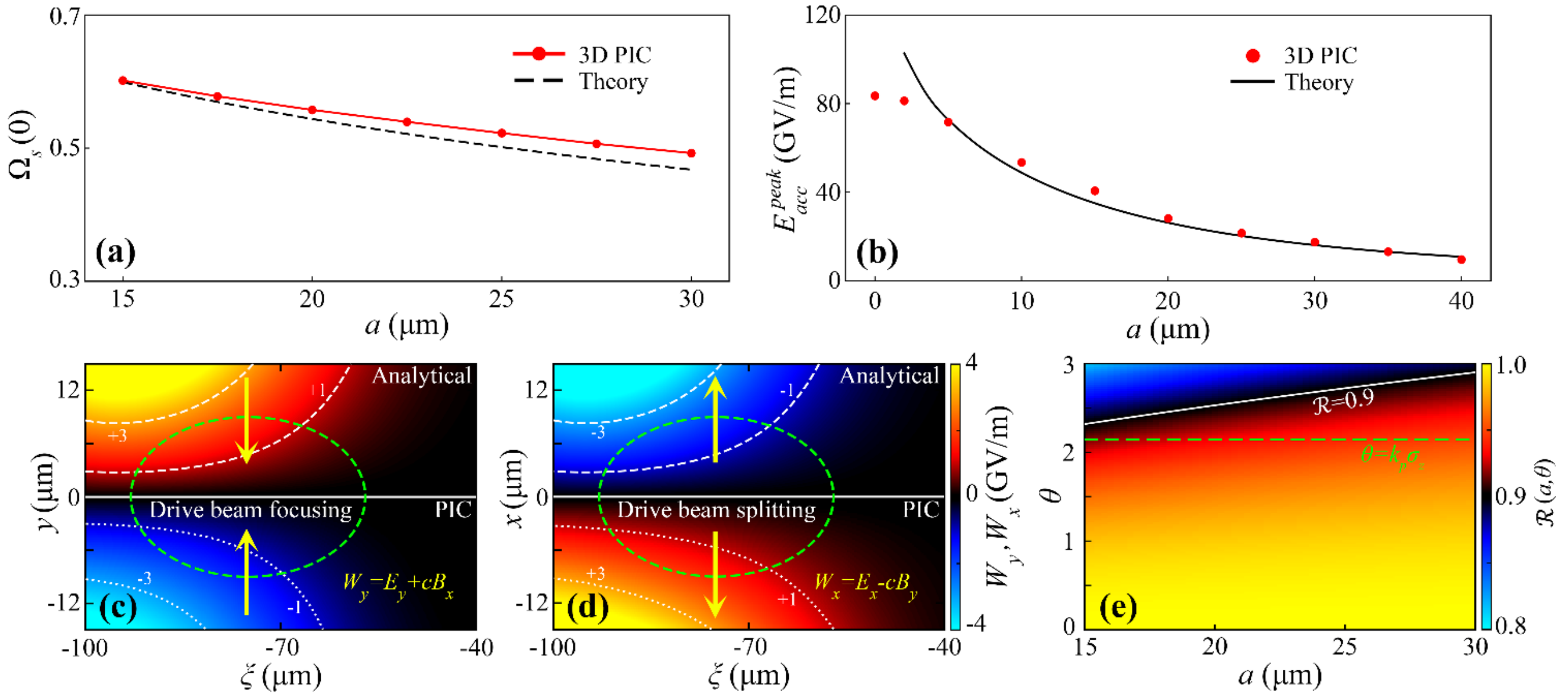

**Fig. 3.** Comparison between theoretical predictions and 3D PIC simulations for slit-plasma wakefields. (a) Eigenfrequency $\Omega_s(0)$ and (b) peak on-axis positron-accelerating field versus slit half-width $a$. (c, d) Transverse fields $W_y$ and $W_x$ at the case $a = 25\mu\text{m}$, with the theoretical prediction shown in the upper part and PIC results in the lower part. Green contours mark the initial $1\sigma$ driver contours. Yellow arrows indicate the force directions on the driver beam. (e) Kernel ratio $\mathcal{R}(a, \theta)$.

The transverse field follows from the Panofsky-Wenzel relation $\partial_\xi \vec{W}_\perp = \nabla_\perp E_z$. Near the axis, $\Phi_{k_y} \simeq 1 + k_y^2(x^2 - y^2)/2$, with $k_y|x|, k_y|y| \ll 1$ for the spectral components contributing appreciably. The constant term has no transverse gradient, whereas the quadratic term generates the transverse structure $x\hat{x} - y\hat{y}$. For the centered driver, integrating the Panofsky-Wenzel relation with $\vec{W}_\perp \to 0$ as $\xi \to +\infty$ gives

$$\vec{W}_\perp(x, y, \xi) \simeq g_s(\xi)(x\hat{x} - y\hat{y}),$$

$$g_s(\xi) = -\int_\xi^\infty \lambda(\xi') K_{slit}\left(\frac{\xi' - \xi}{c}\right) d\xi', \tag{5}$$

$$K_{slit}(\Delta\tau) = -\frac{c}{\pi\omega_p}\int_0^\infty k_y^2\, \kappa_s(k_y) I(k_y) \frac{\sin[\Omega_s(k_y)\omega_p\Delta\tau]}{\Omega_s(k_y)} dk_y.$$

Here $K_{slit}$ is the transverse wake kernel for a source slice, and $g_s(\xi) = \partial_x W_x|_{x=y=0} = -\partial_y W_y|_{x=y=0}$ is the near-axis gradient of the bunch-driven field. The delay $\Delta\tau = (\xi' - \xi)/c$ is measured between the source and observation points. For the driver phases considered here, $g_s < 0$, so the transverse force $-e\vec{W}_\perp$ experienced by the electron beam pushes it outward along $x$ and focuses it toward $y = 0$. Note that the quadrupole-like field is an inherent property of the slit eigenfunction and persists for a symmetric driver. Figures 3(c, d) compare Eq. (5) with simulation results for the case $a = 25\mu\text{m}$, using the parameters of Fig. 2. The analytical model reproduces the field distribution obtained from PIC simulations, with the convolution accounting for the cumulative wake

contributions from preceding source slices. This anisotropic force triggers driver splitting; subsequently, the ion restoring force limits outward motion near the plasma walls.

Offset-induced deflecting wakes can amplify centroid perturbations and induce beam-breakup instability. In the slit geometry, a displacement along $y$ merely translates the wake owing to translational invariance in that direction. A slit-normal offset $\Delta x$, however, excites the odd-$x$ profile $\Psi_{k_y}(x,y) = \sinh(k_y x)\cos(k_y y)$. Averaging $\Psi_{k_y}$ over the displaced transverse beam distribution gives the excitation weight $I_o(k_y) = I(k_y)\sinh(k_y \Delta x) \simeq I(k_y)k_y\Delta x$, where $I(k_y)$ is the centered driver weight. The expansion requires $k_y|\Delta x| \ll 1$ for the modes contributing appreciably to the wake. Near the axis, $I_o \nabla_\perp \Psi_{k_y} \simeq \Delta x k_y^2 I(k_y)\hat{x}$, giving a transversely uniform deflecting wake. Boundary matching yields the odd-branch eigenfrequency $\Omega_o^2(k_y) = \frac{k_y \tanh(k_y a)}{\sqrt{k_y^2 + k_p^2 + k_y \tanh(k_y a)}}$, which is normalized by $\omega_p$. For the same vacuum-field coefficient, the matched even and odd modes have equal modal energies and hence identical spectral loss factors, $\kappa_o(k_y) = \kappa_s(k_y)$ [34]. The deflection kernel $K_{defl}$ therefore follows from $K_{slit}$ in Eq. (5) by replacing $\Omega_s$ with $\Omega_o$. Convolution with the longitudinal charge distribution gives

$$\vec{W}_{\perp,defl}(x,y,\xi) \simeq \Delta x g_o(\xi)\hat{x}, \tag{6}$$

where $g_o(\xi) = -\int_\xi^\infty \lambda(\xi')K_{defl}\left(\frac{\xi'-\xi}{c}\right)d\xi'$ is the deflecting-field response per unit driver offset; the deflecting wake vanishes for a centered driver. We define the dimensionless source-observation delay as $\theta = \omega_p \Delta\tau = k_p(\xi' - \xi)$ and the kernel ratio as

$$\mathcal{R}(a,\theta) \equiv \left|\frac{K_{slit}(\theta/\omega_p)}{K_{defl}(\theta/\omega_p)}\right|. \tag{7}$$

Figure 3(e) evaluates $\mathcal{R}$ from the complete spectral integrals. For $15 \le a \le 30\mu\text{m}$, $0.9 < \mathcal{R} < 1$ for source-observation separations up to one rms bunch length, with $0 < \theta \le k_p\sigma_z = 2.142$; $\mathcal{R} \to 1$ as $\theta \to 0$. For each source slice, the ratio of the even-branch splitting wake at $x \sim \sigma_x$ to the offset-induced odd-branch deflecting wake is $\left|\frac{W_\perp^{split}}{W_\perp^{defl}}\right| = \frac{\sigma_x}{|\Delta x|}\mathcal{R} \simeq \frac{\sigma_x}{|\Delta x|}$. Thus, for $|\Delta x| \ll \sigma_x$, the internal splitting effect dominates the beam dynamics.

With the field structure shown above, we now investigate the performance of positron acceleration. A positron bunch is injected along with the driving beam into the slit plasma channel, as shown in Fig. 4. The witness beam has a Gaussian profile of $\sigma_z = 2\mu\text{m}$ and $\sigma_x = \sigma_y = 1\mu\text{m}$, with 20 pC charge, 100 MeV mean energy, and 0.43% rms energy spread, which is injected into the acceleration phase behind the plasma electron crossing trajectory, specifically centered at $\xi = -168.5\mu\text{m}$. Figure 4(a) shows the energy-trajectory evolution of randomly sampled positrons. Throughout the acceleration process, the positron beam sustains near 100% trapping efficiency with negligible particle loss, while its mean energy exhibits steady linear growth. The transverse focusing field covers the entire volume of the second bubble bucket, as displayed in Fig. 4(b). Due to the focusing field, the positron beam can maintain a highly concentrated density distribution even after traveling 1.5 cm. As shown in Fig. 4(c), transverse field slices at three distinct longitudinal positions [marked in Fig. 4(d)] across the witness beam indicate that the entire positron beam remains tightly enclosed by the focusing fields. Therefore, positrons are effectively confined within the plasma slit region, achieving stable acceleration.

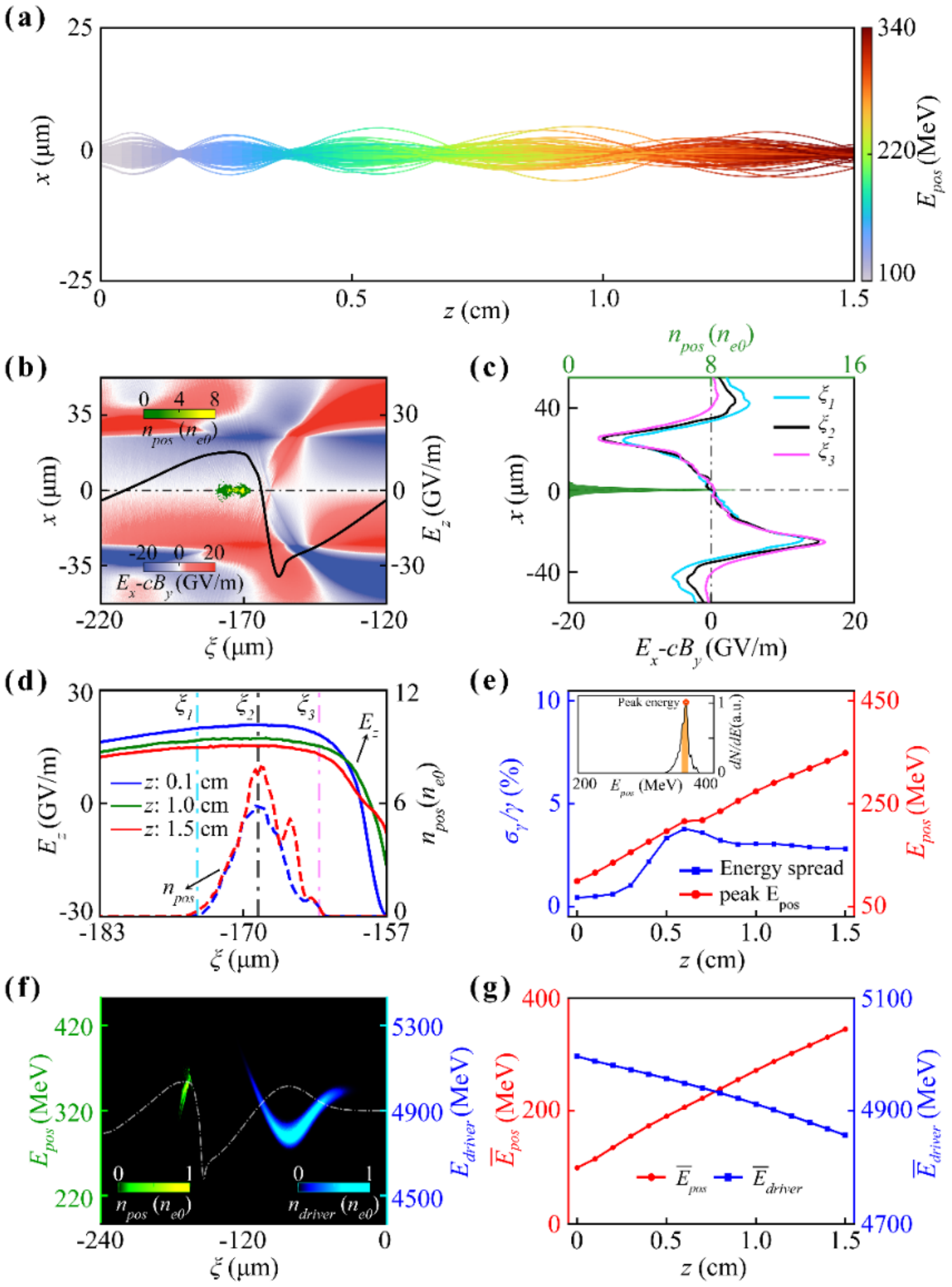

**Fig. 4.** (a) The energy-trajectory evolution of randomly sampled positrons. (b) Transverse field distribution at the $(\xi - x)$ plane, where black curve plots the acceleration field along the propagation axis. (c) Transverse focusing field slices at three different positions: $\xi_1$, $\xi_2$ and $\xi_3$ marked in (d). The green shading displays the final positron density profile in the $x$-direction. (d) Longitudinal acceleration fields at $z = 0.3$cm, 1cm and 1.5cm. The dashed lines plot the longitudinal positron density profile. (e) Evolution of the energy spread and the peak energy of the positron beam as a function of propagation distance. The inset displays the final positron energy spectrum. (f) Longitudinal phase-space distributions of the drive (blue) and witness (green) beams. (g) The average energy evolution of the electron beam and positron beam versus interaction distances.

Figure 4(d) displays the spatial overlap between the on-axis longitudinal accelerating field and the positron density distribution. By loading the positron beam into the appropriate phase region, it can undergo relatively flat acceleration gradients. From the onset of the cavity formation, most of the positrons are stably phase-locked into the accelerating phase. Figure 4(e) illustrates the evolution of the energy spread and peak energy, where the inset shows the final energy spectrum. Thanks to the stabilized longitudinal wakefield, the energy spread can be well constrained below 2.8%, forming a distinct quasi-monoenergetic peak. The positron beam maintains a highly concentrated state in the phase-space distribution [Fig. 4(f)]. The resulting positron energy shows a linear growth trend with the propagation distance, as shown in Fig. 4(g). Optimization can be achieved by tuning the beam and plasma parameters, thereby further improving the final beam quality and energy efficiency.

In real experiments, there might be a density transition region at the interface between vacuum and plasma. To examine this effect, we adopt a super-Gaussian density profile $n_{pe}(x) = n_{e0} \exp\left[-\left(\frac{x_1-|x|}{x_0}\right)^4\right]$ at the region $|x| \leq x_1$ and $n_{pe} = n_{e0}$ for $|x| > x_1$, where $x_0 = 4$μm, and $x_1 = 29$μm, as shown in Fig. 5(a). The obtained results are consistent with those presented above with a step density distribution. Figure 5(b) shows that the positron energy increases almost linearly from the initial 100 MeV to about 770 MeV over a 5 cm distance, with an average acceleration gradient of about 13.4 GV/m. The positron-trapping efficiency remains nearly 100%, with a moderate energy spread of 2.4%. As the positron beam experiences transverse focusing forces throughout acceleration, its emittance is well-preserved, exhibiting negligible growth along the *x*-direction, as shown in Fig. 5(c). During the initial ~1 cm propagation, the evolving driver beam gives rise to variable *y*-direction focusing forces due to the absence of well-defined plasma boundaries. This leads to emittance growth of the positron bunch along the *y*-direction. When the beam driver reaches a quasi-steady state, the positron emittance in the *y*-direction tends to saturate at approximately 20 mm·mrad. If the positron bunch is injected after the beam driver becomes stable, the growth of this emittance can be significantly suppressed, as shown in the inset of Fig. 5(c). Taken together, these features highlight the advantages of our scheme for high-efficiency positron acceleration.

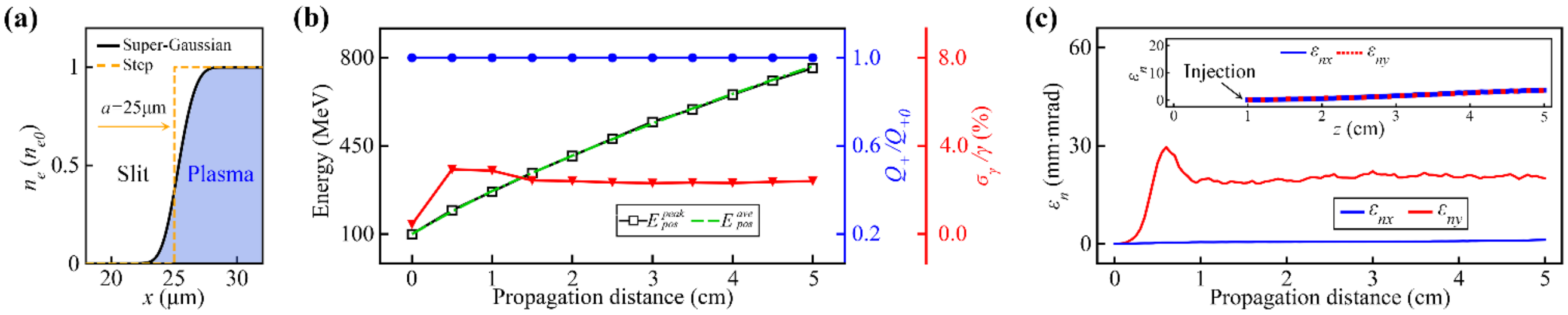

**Fig. 5.** (a) The density profile of the plasma channel. (b) Evolution of the trapping efficiency (blue circles), peak energy (black squares), average energy (green line), and energy spread (red triangles) versus propagation distance. (c) Evolution of the positron beam emittance along the *x*-direction (blue line) and *y*-direction (red line) versus propagation distance. The inset shows the corresponding emittances for the positrons injected after the driver-beam splitting.

In conclusion, we have proposed and numerically demonstrated the SPWA scheme for efficient positron acceleration. In this configuration, plasma electrons flow into the slit channel and generate tailored quasi-static wakefield structures that are capable of providing longitudinal acceleration and transverse focusing simultaneously for trailing positrons. Our approach can overcome the defocusing effect inherent in conventional beam-driven wakefields and achieve a nearly 100% positron trapping efficiency, along with low emittance growth and narrow energy spread. Moreover, transverse beam instabilities can be effectively suppressed, thereby offering high tolerance to beam misalignment. The scheme may open up new opportunities for future compact positron accelerators and a variety of cutting-edge applications.

## *End Matter: The robustness test of the scheme*

To further demonstrate the robustness of the proposed scheme, we apply an initial 1μm offset to the driving beam in the $x$-direction for both the slit channel and the hollow channel configurations for comparison, as shown in Fig. 6. For each case, two types of the beam driver with either circular or elliptic transverse profiles are taken with $(\sigma_x, \sigma_y) = (8, 8)$ and $(12, 6)$ μm, respectively. Here, both the slit half-width and the hollow-channel radius are set to 20μm, and both channels have a 3cm length. All other beam and plasma parameters remain the same as those presented in Fig. 2. In typical hollow channels, when a beam misalignment occurs, the dipole-like fields usually dominate transverse dynamics and continuously deflect the beam in one direction. Therefore, extremely high alignment accuracy is required to achieve positron acceleration. Recent studies suggest that, compared with a symmetric Gaussian beam, a transversely elliptical beam is, to a certain extent, helpful in mitigating such transverse instabilities [23]. Nevertheless, even with such an optimized specially-shaped beam, it is still unable to effectively suppress the beam breakup instability when the transverse offset reaches 1μm. After a 3cm interaction distance, the beam undergoes a significant transverse deviation, seriously distorting the wake structure and eliminating the favorable field required for efficient positron acceleration, as shown in Figs. 6(g)-6(*l*).

In contrast, our scheme employs a routinely accessible Gaussian beam to establish a robust and stable wakefield structure suitable for positron acceleration, as illustrated in Figs. 6(a)-6(f). This simplicity greatly lowers experimental barriers and facilitates near-term experimental implementation. More importantly, it has a higher tolerance for beam misalignment. Under the same 1μm offset condition, the overall evolution of the driver beam and plasma wake remains basically consistent with the on-axis case. Meanwhile, the focusing field is well preserved, and the effective acceleration region for positrons can be stably maintained. Compared with the hollow-channel cases, positrons obtained with our scheme exhibit smaller beam offset and lower emittance, as presented in Figs. 6(m1), 6(m2), 6(n1), and 6(n2). Additionally, for a fixed $|\Delta x|$, a larger $\sigma_x$ of an elliptical beam further enhances the splitting response in the slit-channel configuration, thereby further improving the positron acceleration performance. Benefiting from this unique advantage, our scheme offers superior beam-misalignment tolerance and high structural robustness, making it promising for high-efficiency positron acceleration.

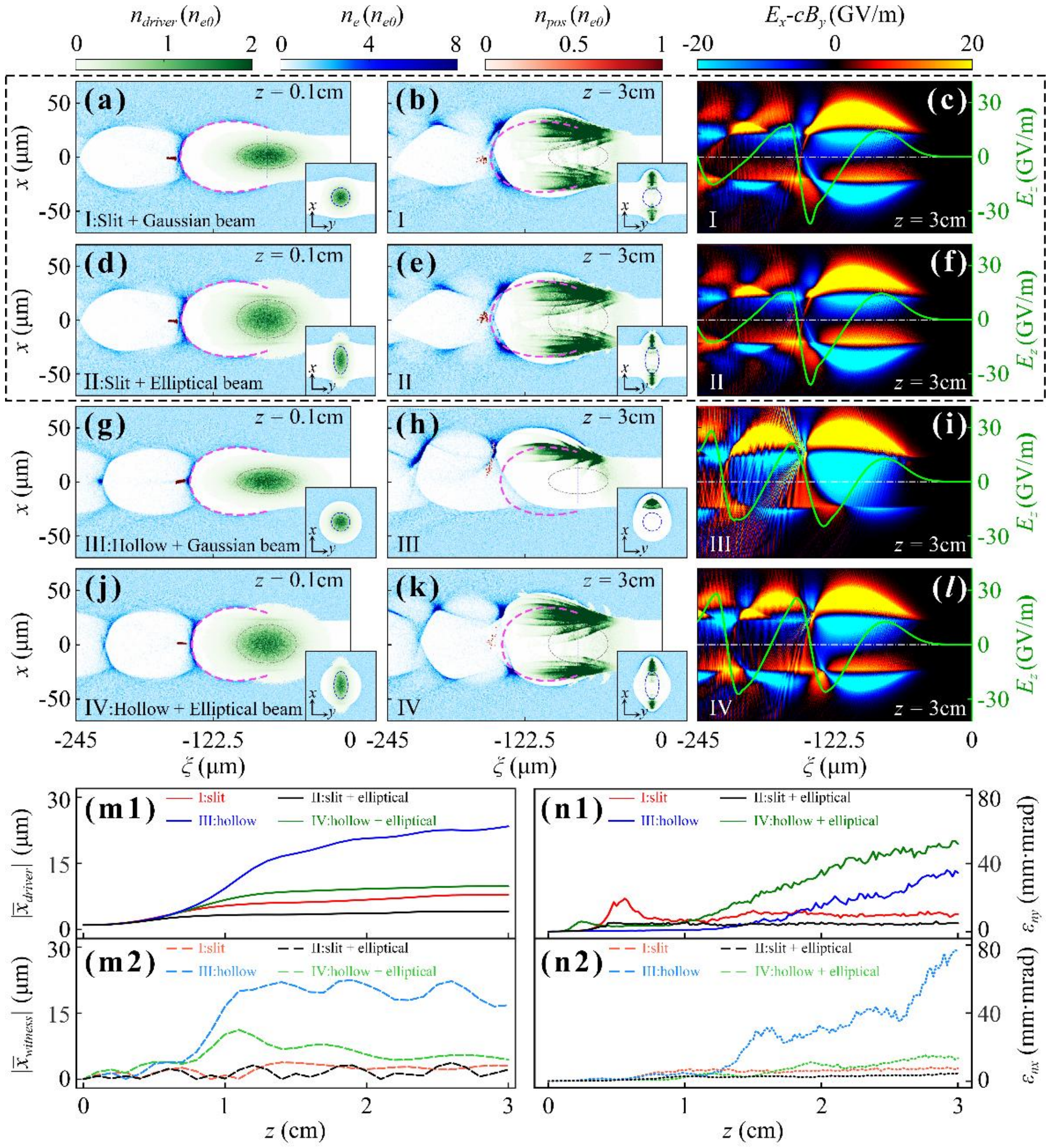

**Fig. 6.** Comparison of the evolution of plasma wakes driven by an electron beam in a slit channel and a hollow channel over a propagation distance of 3cm, where the beam has an initial offset 1μm in the $x$-direction. Four different interaction configurations are shown: (a-c) a Gaussian beam (I) and (d-f) an elliptical beam (II) in a slit-plasma channel; (g-i) a Gaussian beam (III) and (j-$l$) an elliptical beam (IV) in a hollow channel. Among them, the insets show their transverse density profiles at the positions marked by the blue dashed lines. Evolution of the average beam offset $|\bar{x}|$ for (m1) the drive beam and (m2) the witness beam, and the positron beam emittance versus interaction distances for (n1) $\varepsilon_{ny}$ and (n2) $\varepsilon_{nx}$.

**Acknowledgements:** This work was supported by the National Natural Science Foundation of China (Grant Nos. 12522516 and 12595362). X.L.Z. acknowledges the support from Xiaomi Young Scholars Program.

**Data Availability:** The data that support the findings of this study are available from the corresponding author upon reasonable request.